\documentclass[runningheads]{llncs}
\usepackage[T1]{fontenc}
\usepackage{float}
\usepackage{graphicx}
\usepackage[hidelinks]{hyperref}
\usepackage{multirow}
\usepackage{amsmath}
\usepackage{amsfonts} 
\usepackage{booktabs}
\usepackage{adjustbox}
\usepackage{tabularx,ragged2e}
\usepackage{amsfonts}       
\usepackage{nicefrac}       
\usepackage{microtype}      
\usepackage{xcolor}         
\usepackage{todonotes}
\usepackage{placeins}
\usepackage{notoccite}
\usepackage{orcidlink}
\usepackage{comment}
\usepackage{subcaption}

\usepackage[utf8]{inputenc}
\usepackage[most]{tcolorbox}
\usepackage{enumitem}
\usepackage{varwidth}

\usepackage[dvipsnames]{xcolor}

\usepackage{color}

\begin{document}

\title{Motion2Meaning: A Clinician-Centered Framework for Contestable LLM in Parkinson's Disease Gait Interpretation}

\titlerunning{Motion2Meaning}

\author{Loc Phuc Truong Nguyen\inst{1}\orcidlink{0009-0003-4254-0750} \and
Hung Thanh Do\inst{1}\orcidlink{0009-0005-2025-5080}\and\\
Hung Truong Thanh Nguyen\inst{2}\orcidlink{0000−0002−6750−9536} \and
Hung Cao\inst{2}\orcidlink{0000−0002−0788−4377}}
\authorrunning{L.P.T. Nguyen et al.}
%
\institute{Friedrich-Alexander-Universität Erlangen-Nürnberg, 91054 Erlangen, Germany \and
University of New Brunswick, Fredericton E3B 5A3, Canada
\email{\{loc.pt.nguyen,hung.t.do\}@fau.de, \{hung.ntt,hcao3\}@unb.ca}}

\maketitle              
\begin{abstract}
AI-assisted gait analysis holds promise for improving Parkinson's Disease (PD) care, but current clinical dashboards lack transparency and offer no meaningful way for clinicians to interrogate or contest AI decisions. To address this issue, we present Motion2Meaning, a clinician-centered framework that advances Contestable AI through a tightly integrated interface designed for interpretability, oversight, and procedural recourse. Our approach leverages vertical Ground Reaction Force (vGRF) time-series data from wearable sensors as an objective biomarker of PD motor states. The system comprises three key components: a Gait Data Visualization Interface (GDVI), a one-dimensional Convolutional Neural Network (1D-CNN) that predicts Hoehn \& Yahr severity stages, and a Contestable Interpretation Interface (CII) that combines our novel Cross-Modal Explanation Discrepancy (XMED) safeguard with a contestable Large Language Model (LLM). Our 1D-CNN achieves 89.0\% F1-score on the public PhysioNet gait dataset. XMED successfully identifies model unreliability by detecting a five-fold increase in explanation discrepancies in incorrect predictions (7.45\%) compared to correct ones (1.56\%), while our LLM-powered interface enables clinicians to validate correct predictions and successfully contest a portion of the model's errors. A human-centered evaluation of this contestable interface reveals a crucial trade-off between the LLM's factual grounding and its readability and responsiveness to clinical feedback. This work demonstrates the feasibility of combining wearable sensor analysis with Explainable AI (XAI) and contestable LLMs to create a transparent, auditable system for PD gait interpretation that maintains clinical oversight while leveraging advanced AI capabilities. Our implementation is publicly available at: \href{https://github.com/hungdothanh/motion2meaning}{https://github.com/hungdothanh/motion2meaning}.

\keywords{Parkinson's disease care \and Gait analysis \and  Human-centered contestable AI \and  Contestable large language model \and  Explainable AI}
\end{abstract}

\section{Introduction}
The management of chronic neurodegenerative diseases is shifting from episodic evaluations to continuous monitoring with wearable sensors, which provide objective digital biomarkers for earlier intervention and individualized therapy \cite{monje2023co,Daniore2024,info:doi/10.2196/35722}. Parkinson’s Disease (PD), a condition marked by progressive motor impairments \cite{jankovic2008parkinson}, exemplifies this need. Standard clinical tools like the Unified Parkinson’s Disease Rating Scale (UPDRS) \cite{https://doi.org/10.1002/mds.10473} are applied too intermittently and are vulnerable to observer and patient bias \cite{DEDEUSFONTICOBA2019520,AlMahadin2020}. Consequently, they fail to capture daily motor fluctuations, leading to imprecise treatment, heightened fall risk, and reduced quality of life \cite{info:doi/10.2196/35722,https://doi.org/10.1002/mds.30133}.

Although AI models can accurately quantify gait and predict disease severity \cite{10.3389/fneur.2024.1472956,Navita2025}, their clinical translation is stalled by a critical ``last-mile problem.'' Current dashboards present outputs like Hoehn \& Yahr stages \cite{hoehn1967parkinsonism} as opaque, static scores, preventing clinicians from scrutinizing or overriding predictions that conflict with their expertise. This opacity undermines trust and the principles of evidence-based medicine. While Explainable AI (XAI) offers partial solutions like saliency maps \cite{lapuschkin2016lrp,NGUYEN2025102782,ijcai2024p1025}, these are typically one-way communications that fail to support the dialogic nature of clinical reasoning \cite{congait,nguyen2025heart2mind}. The crucial gap is not merely a lack of transparency but the absence of mechanisms for procedural recourse, enabling clinicians to actively contest and amend AI-driven decisions.

To address this gap, we draw on the principles of Contestable AI (CAI). CAI extends beyond explanation by embedding structures for dialogue, challenge, and justification within system design \cite{congait,nguyen2025heart2mind}. This approach aligns with regulatory demands for human oversight (GDPR \cite{eu_gdpr_2016_misc}, EU AI Act \cite{eu2024ai}). A contestable system allows users not only to understand a decision but also to dispute it with domain expertise, ensuring that such challenges are recorded, processed, and capable of influencing the final outcome. This study seeks to apply these principles in a clinician-centered interface for PD care. The key contributions are as follows:
\begin{itemize}
    \item We design and implement \textit{Motion2Meaning}, a novel clinician-centered framework that unifies three core components: a deep learning (DL) diagnostic model, a dual-modality explainability module, and an LLM-driven interaction layer within a single human-in-the-loop interface.

    \item We implement a \textit{1D-CNN architecture} that performs end-to-end classification of Hoehn \& Yahr severity from raw vGRF time-series data. This model outputs a probability distribution over the four discrete severity stages.

    \item We introduce \textit{Cross-Modal Explanation Discrepancy} (XMED), a novel XAI technique to automatically flag unreliable predictions. XMED operates on the principle that trustworthy predictions should have stable explanations across different methods. It quantifies the divergence between a gradient-based explanation (Grad-CAM \cite{selvaraju2017grad}) and a backpropagation-based one (LRP \cite{lapuschkin2016lrp}). A high divergence score signifies inconsistent model reasoning, which automatically flags the prediction for mandatory clinical review.

    \item We develop a \textit{contestable interaction system} powered by a Large Language Model (LLM) that uses a structured ``Contest \& Justify'' workflow. The LLM synthesizes the CNN's prediction, XAI-identified salient features, and the clinician's specific challenge to generate clinically-grounded textual justifications. These justifications form the basis for a transparent, evidence-based dialogue between the clinician and the AI.
\end{itemize}

\section{Background and Related Work}
This work is situated at the confluence of two research domains. We first review advances in AI for sensor-based PD gait analysis, where progress in predictive accuracy has often come at the cost of clinical interpretability. We then connect this gap to the broader evolution of human-centered AI in healthcare, which argues for moving beyond passive Explainable AI (XAI) toward the more interactive and legally robust principles of Contestable AI (CAI).

\subsection{AI for Sensor-Based Parkinson's Gait Analysis}
For decades, gait analysis has been central to movement disorder research, but it was traditionally limited to specialized motion capture laboratories. Wearable Inertial Measurement Units (IMUs) have transformed this field by enabling continuous, high-resolution data collection in natural environments \cite{10.3389/fnhum.2022.768575,MH80258,renggli2020wearable,prisco2024validity}. This is vital for PD care, where gait serves as a rich digital biomarker. Neurodegeneration of dopaminergic pathways in the basal ganglia disrupts movement automaticity, producing measurable deficits in stride length, speed, cadence, turning velocity, and asymmetry. IMUs are also well-suited to detect episodic phenomena such as Freezing of Gait (FOG) and medication-related fluctuations that are difficult to capture in clinic visits \cite{10.3389/fnhum.2022.768575,Park2024}. Early computational methods relied on handcrafted biomechanical features derived from statistical, spectral, and non-linear analyses, which were classified using models such as Support Vector Machines and Random Forests \cite{s21082727}. Although interpretable, these approaches were constrained by their dependence on expert-driven feature design and limited ability to capture complex pathological patterns. More recently, end-to-end DL has emerged, with one-dimensional Convolutional Neural Networks effective for local spatio-temporal motifs, and Recurrent Neural Networks or Transformers capturing long-range dependencies \cite{El_Maachi_2020,naimi20231dconvolutionaltransformerparkinsondisease,rashnu2024integrativedeeplearningframework}. Despite SOTA performance, their opacity poses a major barrier to clinical adoption. This accuracy–interpretability trade-off erodes trust, as clinicians are reluctant to rely on opaque predictions that cannot be examined against their expertise \cite{congait,nguyen2025heart2mind}. The problem is compounded by dashboards that act only as data presenters, showing parameters or outputs without revealing model reasoning or enabling clinician input or correction \cite{hoffman2024comprehensive,congait}.

\subsection{From Explainable to Contestable AI in Healthcare}
\begin{figure}[t!]
    \centering
    \includegraphics[width=0.85\linewidth]{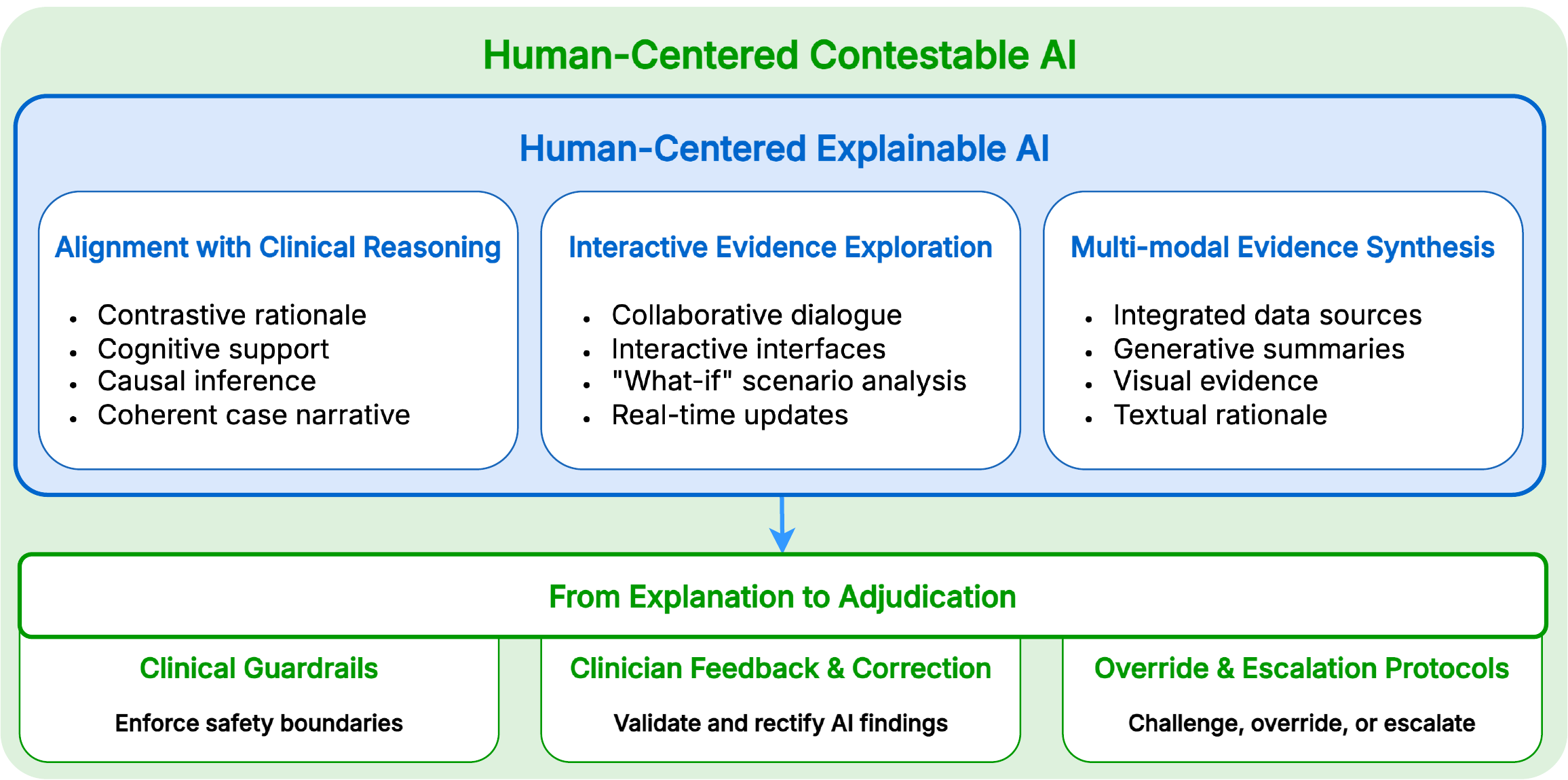}
    \caption{Progression of human-centered XAI toward CAI.}
    \label{fig:xaicai}
\end{figure}

Bridging the gap between high-performance AI and clinical use requires human-centered socio-technical systems that are transparent, interpretable, and trustworthy. This effort began with XAI and is now advancing toward CAI, as illustrated in Figure~\ref{fig:xaicai} \cite{nguyen2025heart2mind}. XAI seeks to make black-box predictions understandable, supporting trust and error detection \cite{10.1007/978-3-032-02813-6_9,ijcai2024p1025,NGUYEN2025102782}. In clinical gait analysis, XAI remains early, though methods from other medical domains provide guidance. Backpropagation-based approaches, such as Saliency Maps and LRP \cite{lapuschkin2016lrp}, highlight critical temporal regions of the input. CAM-based techniques, including Grad-CAM \cite{selvaraju2017grad} and Grad-CAM++ \cite{8354201}, localize discriminative regions linked to predictions. Perturbation-based methods \cite{10.1145/2939672.2939778,petsiuk2018riserandomizedinputsampling,truong_towards_2024} identify influential regions by altering inputs and monitoring changes in output probabilities. More recent work emphasizes interactive explanations, allowing users to test counterfactuals or adjust inputs, and cognitively aligned formats, such as contrastive reasoning or natural language dialogue \cite{congait,nguyen2025heart2mind}. These developments recognize explanation as a social process aimed at shared understanding between humans and AI.

A central challenge in human-AI collaboration is achieving appropriate trust calibration \cite{Romeo2025,li2024overconfident}, the process by which a user develops an accurate mental model of an AI's capabilities to avoid both blind over-trust (automation bias) and reflexive dismissal (algorithm aversion). Conventional XAI, while providing transparency, may not suffice for this task. A compelling but incorrect explanation can actively impair calibration by creating a false sense of security \cite{10.1145/3351095.3372852}. A system designed for effective calibration must therefore go beyond one-way explanations and provide a mechanism for procedural recourse \cite{herrera2025making}. This is the principle of CAI: to create an essential feedback loop where clinicians can act on their calibrated judgments, transforming them from passive observers into active supervisors, enabling expert-driven recourse that forms the foundation of our Motion2Meaning framework, which aims to operationalize contestability in a real-world clinical setting.
\section{Methodology}

\begin{figure}[ht!]
    \centering
    \includegraphics[width=0.85\linewidth]{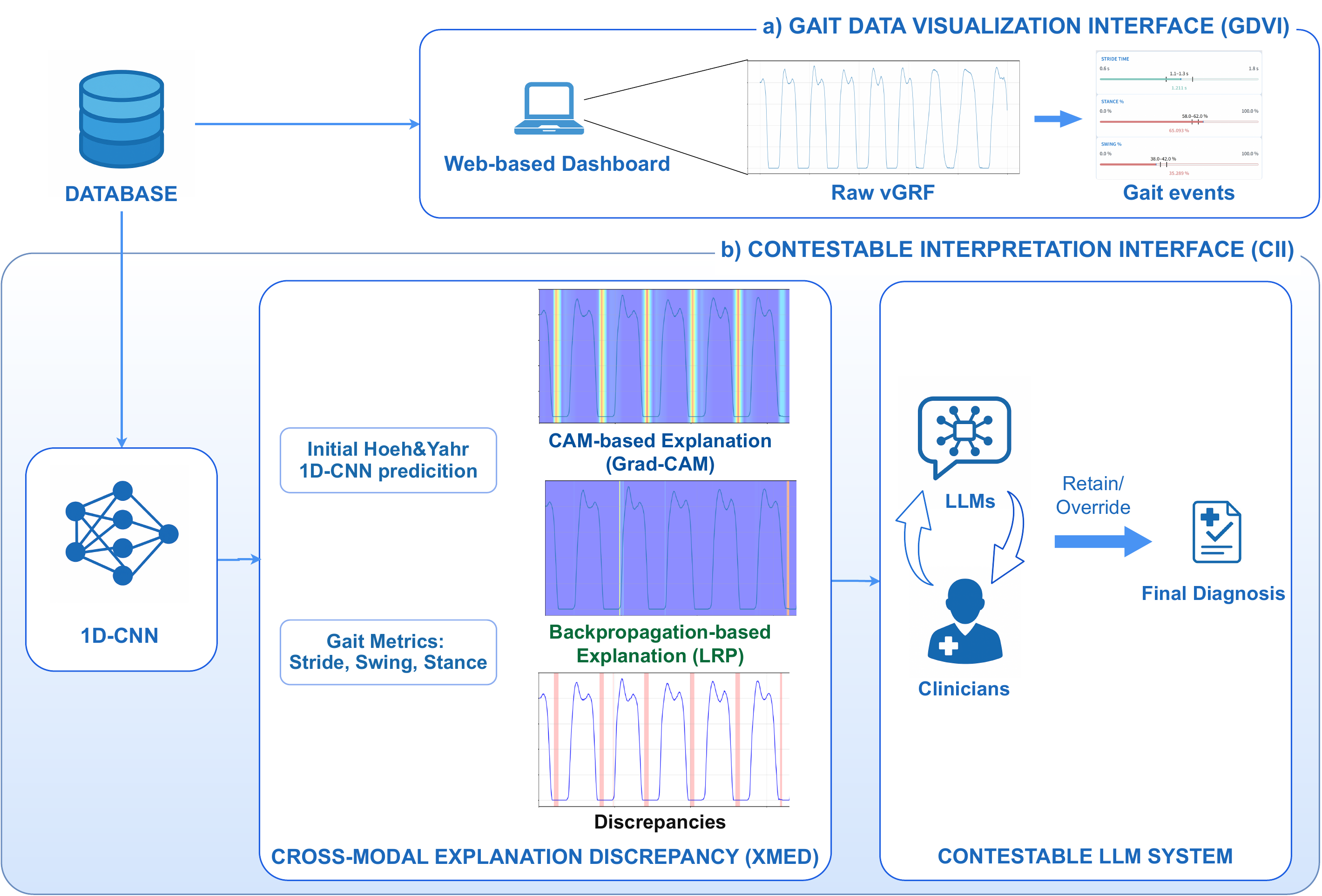}
    \caption{The overview of Motion2Meaning framework: (a) Gait Data Visualization Interface (GDVI), and (b) Contestable Interpretation Interface (CII).}
    \label{fig:overview}
\end{figure}

Figure~\ref{fig:overview} presents an overview of the Motion2Meaning framework, which integrates two core interfaces into an end-to-end system for gait interpretation.

\begin{figure}[ht!]
    \centering
    \includegraphics[width=0.85\linewidth]{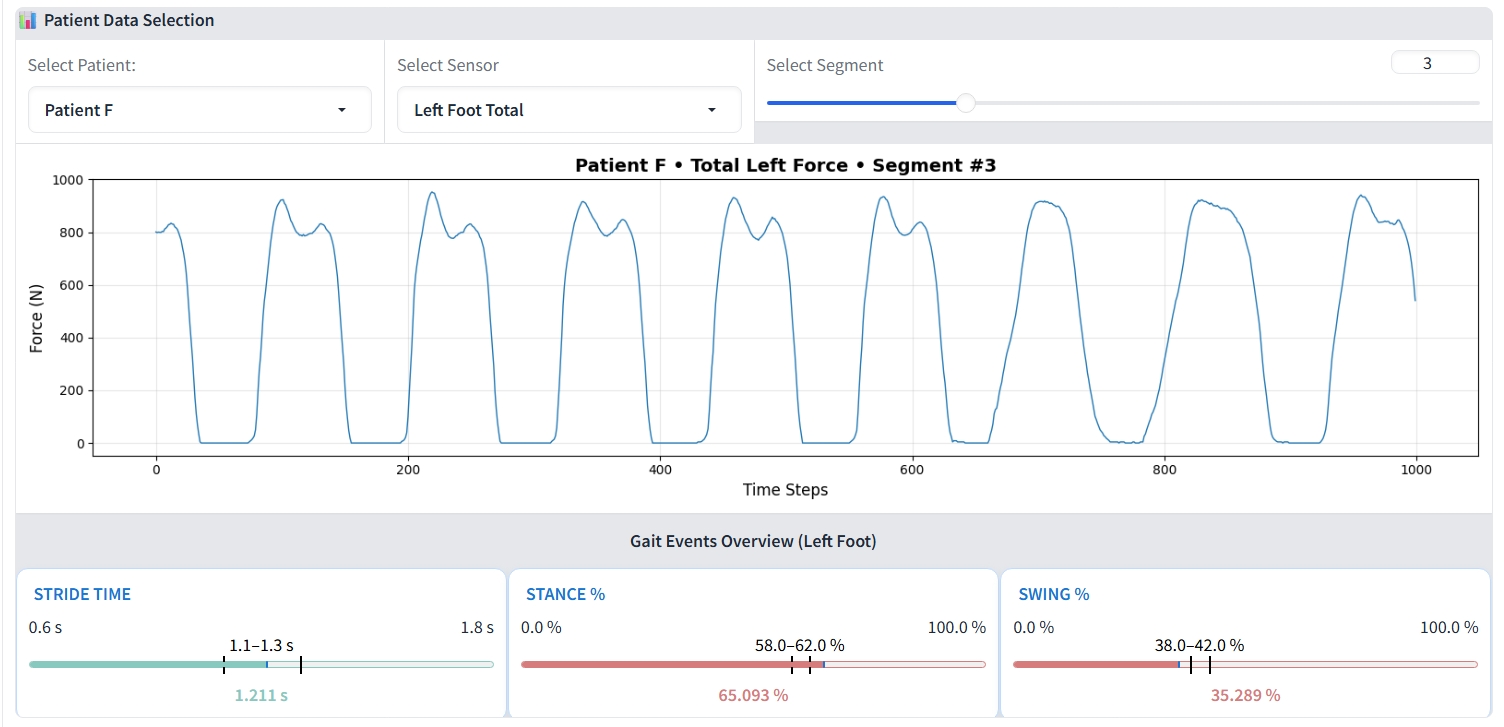}
    \caption{The dashboard overview of Gait Data Visualization Interface (GDVI).}
    \label{fig:gdvi}
\end{figure}

The first component is the \textbf{Gait Data Visualization Interface} (GDVI). It is an interactive web-based tool for exploring gait data from the PhysioNet dataset \cite{goldberger2000physiobank}, as depicted in Figure~\ref{fig:gdvi}. The interface presents raw vGRF signals in 10-second windows, with controls to toggle sensor channels and select time segments for detailed inspection. A complementary summary panel uses color-coded markers to highlight key temporal events, including Stride, Stance, and Swing Time. This design enables clinicians to rapidly compare gait patterns against normative data and visually identify potential anomalies.

The second and also core component of the framework is the \textbf{Contestable Interpretation Interface} (CII), the dashboard where the human-AI dialogue occurs. Its workflow is operationalized through three integrated technical pillars:

\begin{enumerate}

    \item The workflow begins with the \textit{1D-CNN}, our predictive engine designed following \cite{10.3389/fninf.2024.1451529}, which analyzes a patient's gait data to generate an initial classification of the Hoehn \& Yahr severity. This prediction is presented not as a final answer, but as a testable hypothesis for clinical review.

    \item To audit the reliability of this hypothesis, the \textit{Cross-Modal Explanation Discrepancy} (XMED) module performs an automated consistency check on the model's reasoning. It leverages the fundamental differences between Grad-CAM \cite{selvaraju2017grad}, which identifies \textit{where} in the signal the model focuses, and LRP \cite{lapuschkin2016lrp}, which attributes \textit{what} specific data points were most influential. A significant divergence between these two reveals a critical failure mode where the model may correctly identify a clinically relevant temporal region but base its decision on a spurious artifact within it. This quantified "attention-attribution" gap provides a targeted alert for a structurally flawed reasoning process that a single explanation method would miss.

    \item Finally, the \textit{LLM-powered "Contest \& Justify" workflow} enables procedural recourse. When a prediction is challenged, either due to an XMED alert or independent clinical judgment, the clinician registers dissent through a structured typology: \textit{Factual Error} (contesting input data integrity), \textit{Normative Conflict} (flagging contradiction with clinical knowledge), or \textit{Reasoning Flaw} (challenging the XAI's visual evidence). This formal contestation triggers the LLM to synthesize all available evidence and generate a new, contextualized justification. This exchange creates a transparent, collaborative negotiation that culminates in either acceptance or a clinician-driven override, with every step logged in an immutable record to ensure accountability.
\end{enumerate}

\section{Experiment and Results}
To evaluate the effectiveness of our Motion2Meaning framework, we conducted a comprehensive, multi-stage investigation. Our evaluation was designed to answer three central questions: (1) \textit{What is the baseline predictive performance of our core 1D-CNN model on PD severity classification?}; (2) \textit{Can the XMED method effectively distinguish between reliable and unreliable model predictions?}; (3) \textit{How effectively can LLMs leverage these discrepancy signals to validate correct predictions and contest erroneous ones in a simulated clinical workflow?}

\subsection{Dataset and Experimental Setup}
Our experiments were conducted using the public PhysioNet Gait in PD dataset, which contains vertical Ground Reaction Force (vGRF) signals from 93 individuals with PD and 73 healthy controls \cite{goldberger2000physiobank}. For our deep learning model, we preprocessed the data by segmenting the variable-length recordings into fixed, non-overlapping 1000-frame windows. To create a focused and interpretable attribution space for our XMED safeguard, we used a single, highly informative feature for our analysis: the "Total left force" signal.

The dataset was partitioned into training (70\%), validation (15\%), and test (15\%) sets, using multiple random seeds to ensure robustness. We trained our model using a nested cross-validation strategy on the training data. A 5-fold outer loop assessed model generalization, while a 3-fold inner loop within each fold conducted a grid search to optimize hyperparameters. The final optimal configuration derived from this process is presented in Table~\ref{tab:cnn_hyperparameters}.

\subsection{Human-Centered Evaluation Metrics}
To align with CAI and foreground the framework’s human-centered design, we evaluate Motion2Meaning using human-oriented metrics. We concentrate on LLM-generated textual explanations, which constitute the most direct interface between the AI component and clinicians. Our first two metrics are the Flesch readability tests \cite{kincaid1975derivation,flesch1979write}, which estimate readability via sentence length and lexical complexity. \textbf{Flesch Reading Ease (FRE)} ranges from 1 to 100, with higher values indicating greater accessibility. The \textbf{Flesch-Kincaid Grade Level (FKGL)} estimates the U.S. school grade needed to comprehend a text. For clinician-oriented medical materials (e.g., clinical documentation used in diagnostic and care workflows), typical FRE scores are 50 to 70. These correspond to FKGL 8 to 12 and are appropriate for readers aged approximately 13 to 18 \cite{wu2013applying,challener2025flesch}. The corresponding formulas are given as:

\begin{align*}
\text{FRE} &= 206.835 - 1.015 \left( \frac{\text{total words}}{\text{total sentences}} \right) - 84.6 \left( \frac{\text{total syllables}}{\text{total words}} \right), \\
\text{FKGL} &= - 15.59 + 0.39 \left( \frac{\text{total words}}{\text{total sentences}} \right) + 11.8 \left( \frac{\text{total syllables}}{\text{total words}} \right) .
\end{align*}

\textbf{Clinical Grounding (CG)} evaluates LLM hallucination by quantifying the verifiability of its explanations against available evidence. We compute it by first isolating all numerical values in the model's generated text, and then determining the percentage of those numbers that match the figures provided in the input prompt and data. Given the multiset of numerical values extracted from the LLM's generated explanation, $V_E$, and the multiset of all numerical values provided in the input prompt and data, $V_I$, CG is defined as:

\begin{align*}
\text{CG} = \frac{100}{|V_E|} \sum_{\nu \in V_E} \mathbb{I}(\nu \in V_I),
\end{align*}
where $\mathbb{I}(\cdot)$ is the indicator function and the score is defined as 100 if $\lvert V_E \lvert = 0$ (i.e., no numerical claims are made). A high score indicates the framework's ability to reduce a clinician's cognitive load and mitigate clinical risk. 

\textbf{Self-Correction Accuracy (SCA)}, measures the direct impact of our contestation system on fixing the baseline model's mistakes. We compute it by first isolating the set of instances $\mathcal{D}_{\text{err}}$ that the baseline model initially misclassified, and then calculating the percentage of these specific errors that the contestable LLMs system successfully overturns to the correct label. Given the final system's prediction $\hat{y}_{\text{final}}(x_i)$ and the true label $y_i$ for an instance $x_i \in \mathcal{D}_{\text{err}}$, SCA is formally defined as:

\begin{align*}
\text{SCA} = \frac{100}{|\mathcal{D}_{\text{err}}|} \sum_{x_i \in \mathcal{D}_{\text{err}}} \mathbb{I}(\hat{y}_{\text{final}}(x_i) = y_i),
\end{align*}
where $\mathbb{I}(\cdot)$ is the indicator function. A high score confirms the framework's self-remediation capacity, enhancing user trust and clinician-AI collaboration.

\subsection{Predictive Model Performance}
Upon evaluation on the unseen test set, our model achieved a robust overall accuracy and weighted F1-score of 0.89, with a detailed breakdown of per-class performance provided in Table~\ref{tab:classification-report}. The model demonstrates excellent performance in identifying the Healthy control group, achieving an F1-score of 0.91. Furthermore, it shows reliable discrimination between the clinically adjacent intermediate severity levels, with balanced F1-scores of 0.87 for Stage 2 and 0.89 for Stage 2.5. The primary challenge was observed in the most advanced category, Stage 3, which recorded a slightly lower recall of 0.84. This reduced sensitivity is likely attributable to the significant class imbalance, as this category contains only 83 samples, which may limit the model's ability to learn its full intra-class variability. These results confirm the model's overall robustness but also highlight a clear direction for future refinement; techniques such as targeted data augmentation or class-balancing loss functions could further improve sensitivity for the more advanced disease stages.

\begin{table}[b!]
\centering
\begin{minipage}{0.43\textwidth}
\centering
\captionof{table}{Optimized hyperparameter configurations for the 1D-CNN model.}
\label{tab:cnn_hyperparameters}
\begin{adjustbox}{width=\textwidth}
\begin{tabular}{l c}
\hline
\textbf{Hyperparameter} & \textbf{Value} \\
\hline
Number of Convolutional Layers & 5 \\
Number of Fully Connected Layers & 3 \\
Activation Function & ReLU \\
Dropout Rate & 0.5 \\
Learning Rate & 0.0003 \\
\hline
\end{tabular}
\end{adjustbox}
\end{minipage}
\hfill
\begin{minipage}{0.56\textwidth}
\centering
\captionof{table}{Classification performance of the 1D-CNN model on the test set.}
\label{tab:classification-report}
\begin{adjustbox}{width=\textwidth}
\begin{tabular}{lcccc}
\hline
\textbf{Class} & \textbf{Precision} & \textbf{Recall} & \textbf{F1-Score} & \textbf{Support} \\
\hline
Healthy   & 0.92 & 0.90 & 0.91 & 278 \\
Stage 2   & 0.85 & 0.89 & 0.87 & 346 \\
Stage 2.5 & 0.90 & 0.88 & 0.89 & 228 \\
Stage 3   & 0.92 & 0.84 & 0.88 & 83 \\
\hline
Accuracy  &      &      & 0.89 & 935 \\
Macro Avg & 0.90 & 0.88 & 0.89 & 935 \\
Weighted Avg & 0.89 & 0.89 & 0.89 & 935 \\
\hline
\end{tabular}
\end{adjustbox}
\end{minipage}

\end{table}

\subsection{Analysis of the Contestable AI System}
The model's fallibility, particularly in intermediate and advanced disease stages, highlights the need for a human-in-the-loop system to identify and correct errors. We first evaluated our XMED safeguard, which is based on the hypothesis that discrepancies between explanation methods can serve as a proxy for model uncertainty. To test this, we quantified the ``high-discrepancy percentage'' for a set of 30 test cases. The results confirm our hypothesis: misclassified cases exhibited a five-fold higher average discrepancy rate (7.56\%) compared to correct predictions (1.45\%). This demonstrates that attributional inconsistency is a reliable signal of model unreliability.

\begin{figure}[t!]
    \centering
    \includegraphics[width=0.85\linewidth]{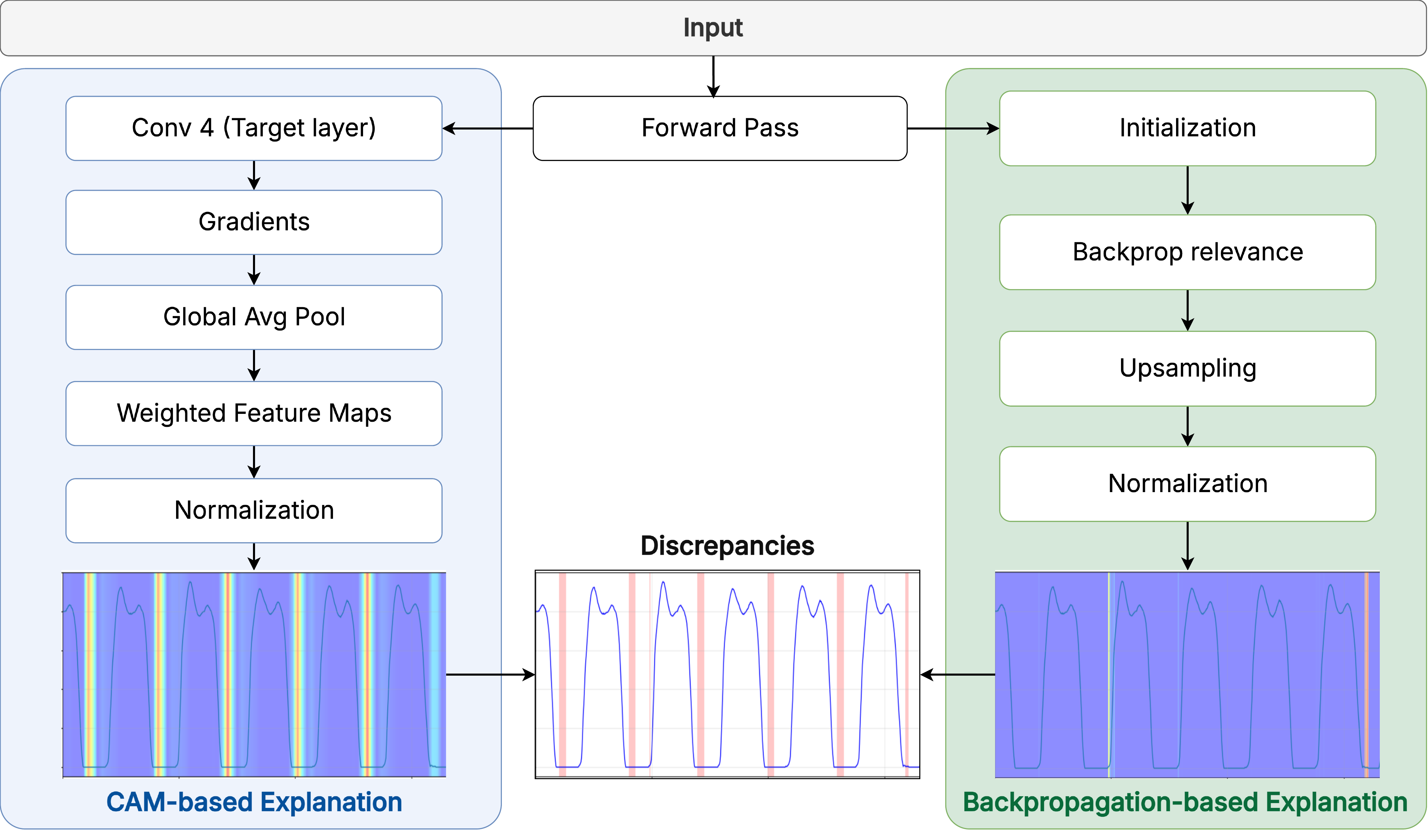}
    \caption{Workflow overview of the XMED. The process compares CAM-based (Grad-CAM) and backpropagation-based (LRP) explanations to quantify model uncertainty. The input undergoes a forward pass to extract activations from the target convolutional layer. Grad-CAM computes weighted feature maps via gradient-based pooling, while LRP propagates relevance scores backward through the network. Both maps are normalized and compared to identify regions of high discrepancy, indicating divergent model explanations.}
    \label{fig:xmed-flowchart}
\end{figure}

To calculate this metric, we generated two normalized explanation maps for each test sample: one using Grad-CAM \cite{selvaraju2017grad} and another using LRP \cite{lapuschkin2016lrp}, as illustrated in Figure~\ref{fig:xmed-flowchart}. We calculated the absolute difference at each timestep, flagging points where it exceeded a threshold of 0.5. These points were then merged into coherent high-discrepancy regions, and the final metric was the fraction of total timesteps falling within these regions.

With the safeguard validated, we investigated whether LLMs could use these signals for adjudication. We tested Llama 4 Scout (17B) \cite{meta2025llama4} and GPT-4o (200B) \cite{hurst2024gpt} with identical prompts (Template~\ref{template:A}) and settings. As summarized in Table~\ref{tab:contestable_llms_cases}, their performance profiles differed significantly. GPT-4o adopted a more conservative and reliable approach, correctly retaining all 24 correct predictions it reviewed. In contrast, Llama 4 was more interventionist; it successfully overturned two of the six incorrect predictions but also incorrectly overturned one correct case, suggesting different underlying reasoning processes that warrant further case-level analysis.

\newtcolorbox[auto counter]{prompttemplate}[1][]{
  enhanced,
  fonttitle=\scshape,
  #1
}
\definecolor{ForestGreen}{RGB}{34,139,34}
\begin{figure*}[t!]
\begin{prompttemplate}[label=template:A,
  title={Template \thetcbcounter: Prompt template for Contestable Gait Analysis}
]
\scriptsize
\textbf{SYSTEM MESSAGE:} 
You are a helpful clinical decision support AI for Parkinson's disease diagnosis using gait analysis. Always:
\begin{enumerate}
    \item Think step-by-step before responding.
    \item Justify your initial assessment and interpretation of gait metrics, referencing clinical guidelines or evidence when possible.
    \item When finalization request is queried, you must finalize the decision (only answer: ``Healthy'', ``Stage 2'', ``Stage 2.5'', or ``Stage 3'') but you may overturn your prior assessment if, after reviewing all evidence, you are confident a different answer is correct.
    \item Using your clinical analysis and justification, identify the potential reasons for any change in the final decision (e.g., specific gait abnormalities, asymmetries, variability, etc.) or in case of no change, justify why the initial assessment was correct. Then explain how these factors contribute to the final severity.
    \item Provide accurate, current information using clinical gait analysis guidelines.
    \item Cross-validate findings with multiple sources.
    \item Reference sources for non-standard conclusions.
    \item Maintain clarity with concise and straightforward responses.
\end{enumerate}
\textbf{USER MESSAGE:}
\begin{enumerate}
    \item \textbf{Prediction:} \{\textcolor{ForestGreen}{class}\} (Confidence: \{\textcolor{ForestGreen}{confidence}\}).
    \item \textbf{Gait metrics}: 
    \{\textcolor{ForestGreen}{mean\_stride\_time}\}, 
    \{\textcolor{ForestGreen}{swing\_percentage}\}, 
    \{\textcolor{ForestGreen}{stance\_percentage}\}
    \item \textbf{XAI analysis output:}
        \begin{itemize}
            \item Discrepancy percentage: \{\textcolor{ForestGreen}{discrepancy\_percentage}\}\%.
            \item Continuous high-discrepancy regions: \{\textcolor{ForestGreen}         {[region\_1], [region\_2],...,[region\_n]}\}.
        \end{itemize}
\end{enumerate}
\textit{*Note that the numerical input values (in \textcolor{ForestGreen}{green}) have been pre-calculated by the baseline model and XMED.}
\end{prompttemplate}
\end{figure*}

\subsubsection{Correct Prediction Case}
\begin{figure}[ht!]
    \centering
    \includegraphics[width=0.9\linewidth]{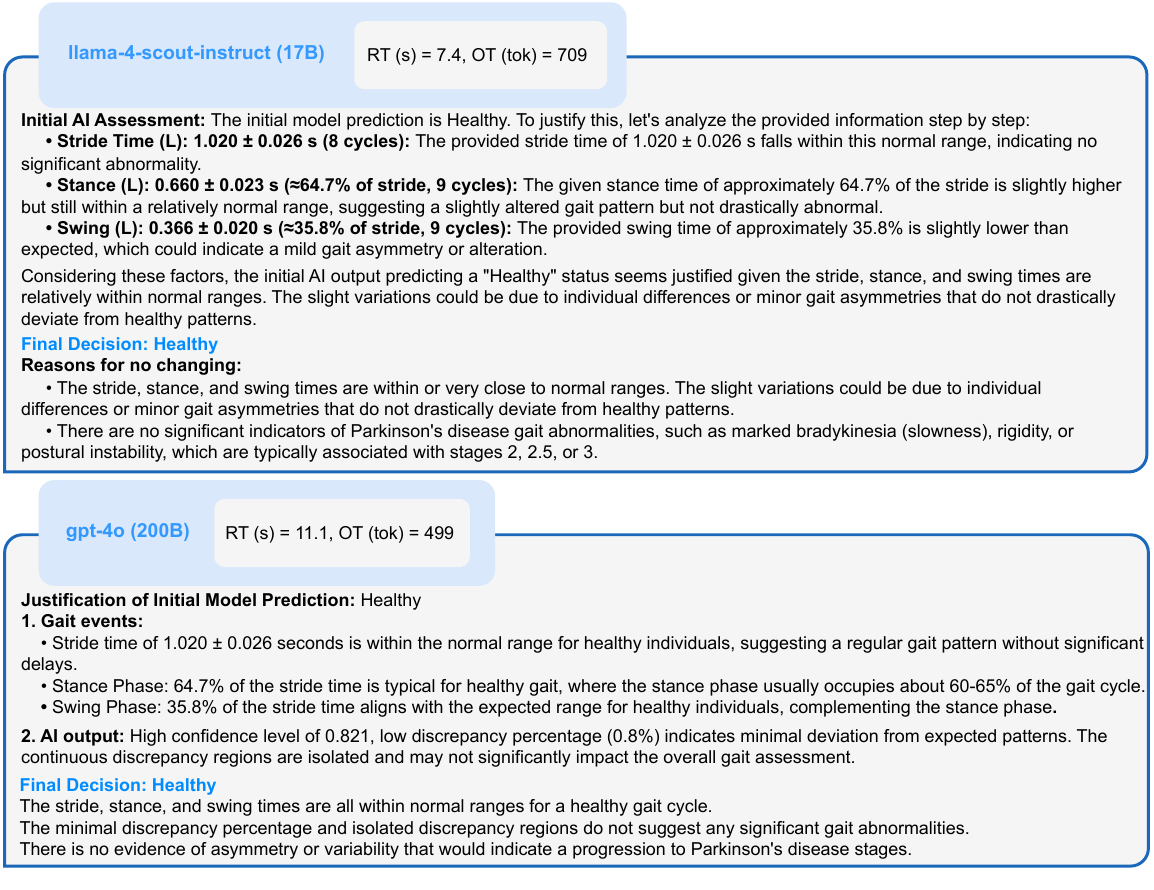}
    \caption{Example responses from contestable LLMs in a Correct Prediction case. The final decision (in \textcolor{NavyBlue}{blue}) indicates the model retains the correct initial prediction of baseline 1D-CNN. Output metrics: response time - RT (s) and number of output token - OT (tok).}
    \label{fig:correct-llm}
\end{figure}

In a representative case of a correct classification (Figure~\ref{fig:correct-xmed}), the 1D-CNN identified a patient as Stage 0 (Healthy) with high confidence (0.821) and a correspondingly low XMED discrepancy of 0.8\%. Both LLMs correctly upheld this initial prediction. As detailed in Figure~\ref{fig:correct-llm}, their justifications were grounded in clinical gait data, with both models observing that gait metrics were ``within normal ranges.'' Llama 4 further specified the ``absence of Parkinsonian gait markers,'' while GPT-4o correctly interpreted the minor discrepancy regions as not clinically significant, demonstrating a nuanced understanding of the XMED signal.

\subsubsection{Incorrect Prediction Case}
\begin{figure}[t!]
    \centering
    \begin{subfigure}[b]{0.44\linewidth}
        \centering
        \includegraphics[width=\linewidth]{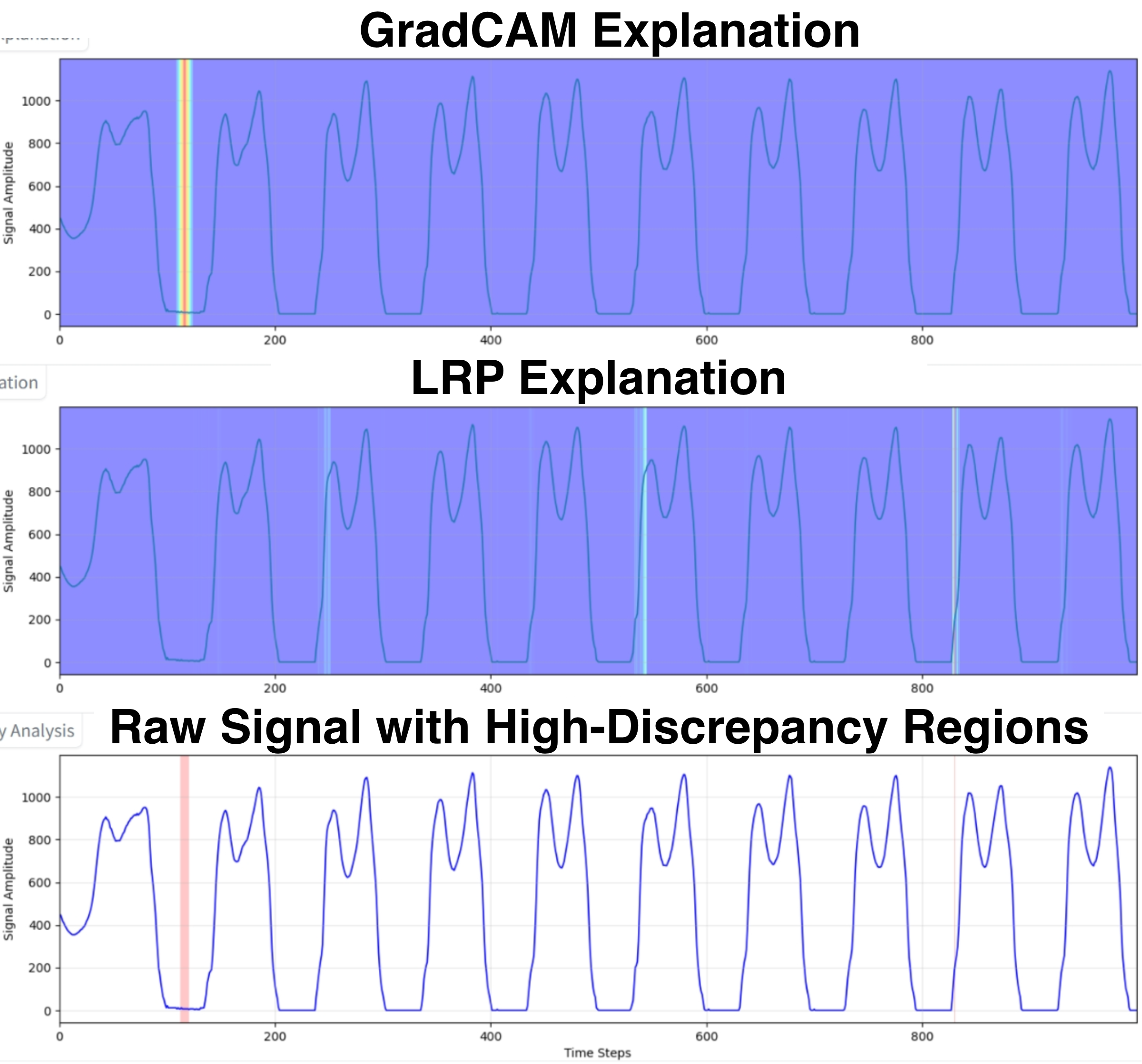}
        \caption{Correct Prediction}
        \label{fig:correct-xmed}
    \end{subfigure}
    \hfill
    \begin{subfigure}[b]{0.44\linewidth}
        \centering
        \includegraphics[width=\linewidth]{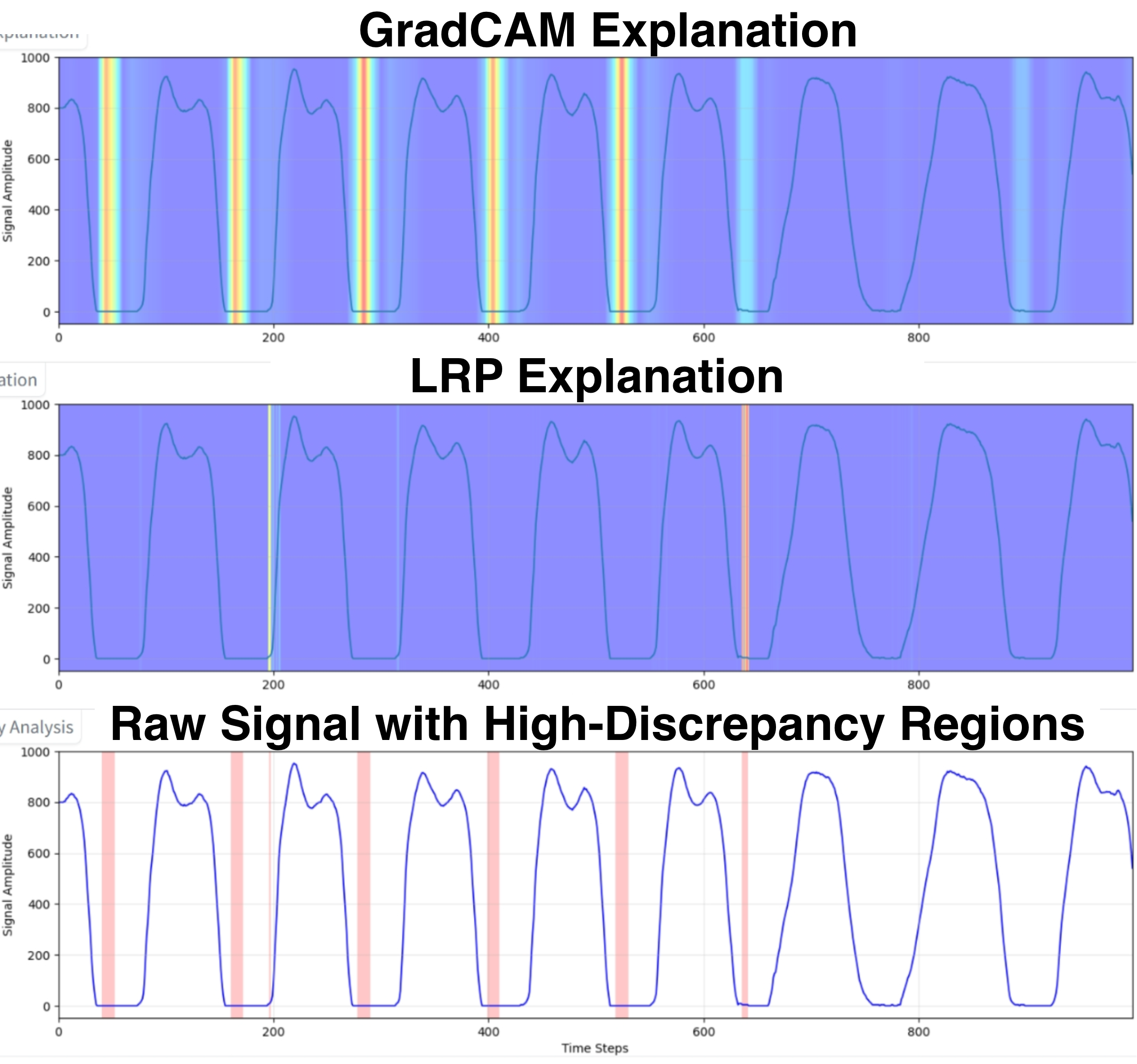}
        \caption{Incorrect Prediction}
        \label{fig:incorrect-xmed}
    \end{subfigure}
    \caption{XMED Visualization for (a) correct and (b) incorrect prediction cases.}
    \label{fig:xmed-comparison}
\end{figure}

\begin{figure}[ht!]
    \centering
    \includegraphics[width=0.9\linewidth]{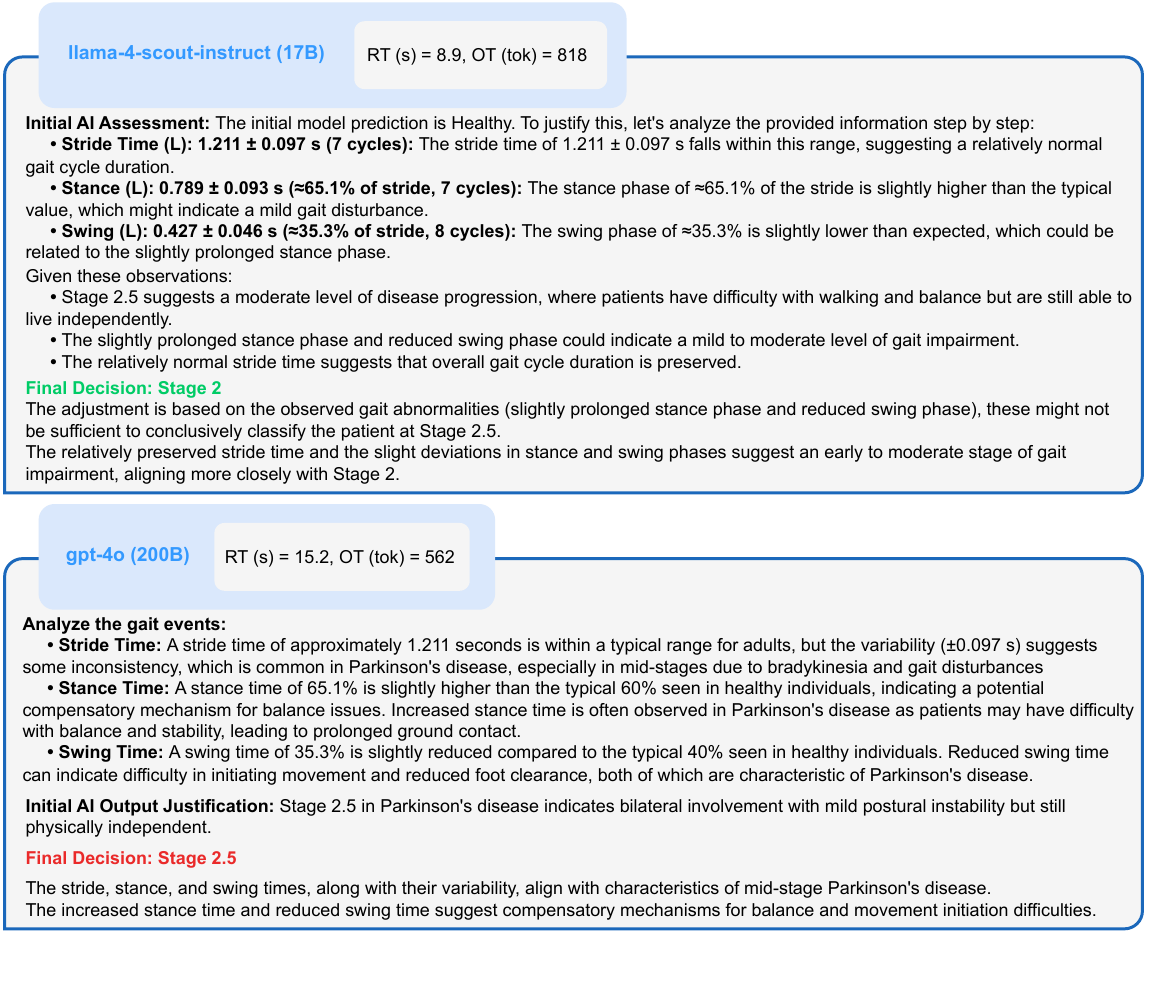}
    \caption{Example responses from contestable LLMs in an Incorrect Prediction case. The final decision (in \textcolor{red}{red}) indicates the model retains the incorrect initial prediction of baseline 1D-CNN, while the final decision (in \textcolor{ForestGreen}{green}) indicates the model overturns it. Output metrics: response time - RT (s) and number of output token - OT (tok).}
    \label{fig:incorrect-llm}
\end{figure}

A more revealing case involved a low-confidence (0.462) misclassification by the 1D-CNN, which was correctly flagged by a high XMED score of 6.5\% (Figure~\ref{fig:incorrect-xmed}). In this instance, the LLMs diverged, as shown in Figure~\ref{fig:incorrect-llm}. Llama 4 successfully overturned the prediction, downgrading it from Stage 2.5 to Stage 2. It correctly reasoned that while the prolonged stance phase (65.1\%) indicated impairment, the abnormality was insufficient to justify the higher severity rating. In contrast, GPT-4o retained the incorrect Stage 2.5 label, focusing on the clinical plausibility of the observed gait changes as compensatory mechanisms. It failed to differentiate the degree of this deviation, highlighting a more risk-averse adjudicative style that avoids overturning a prediction without overwhelming contradictory evidence.

In terms of computational efficiency, we observed a clear trade-off. Llama 4 was consistently faster, with a response time (RT) between 7 and 9 seconds, while GPT-4o was slower, taking between 11 and 15 seconds. However, GPT-4o produced more concise and direct justifications, with an output token (OT) count of approximately 499-562, compared to Llama 4's more verbose outputs of 709-818 tokens. This suggests that while the smaller model offers lower latency, the larger model provides superior adjudication quality and more clinically-grounded reasoning.

\begin{table}[t!]
    \centering
    \caption{Number of classification cases by contestable LLMs with XMED support.
    The arrows ($\uparrow$/$\downarrow$) indicate higher/lower is better.}
    \begin{tabularx}{\linewidth}{l *{4}{>{\centering\arraybackslash}X}}
    \toprule
    \textbf{Model} & 
    \textcolor{NavyBlue}{Retain Correct $\uparrow$} & 
    \textcolor{BrickRed}{Retain Incorrect $\downarrow$} & 
    \textcolor{BrickRed}{Overturn Correct $\downarrow$} & 
    \textcolor{ForestGreen}{Overturn Incorrect $\uparrow$} \\
    \midrule
    llama-4-scout-instruct (17B) & 23 & 4 & 1 & 2 \\
    gpt-4o (200B)  & 24 & 5 & 0 & 1 \\
    \bottomrule
    \end{tabularx}
    \label{tab:contestable_llms_cases}
\end{table}


\subsection{Human-centered Evaluation Results}
To assess the framework's human-centered design in line with CAI principles, we evaluated Motion2Meaning on four key metrics. For a consistent and robust evaluation, these metrics were computed and averaged across the 30 test cases from the previous experiment, with the results summarized in Table \ref{tab:human-metrics}.

\subsubsection{Flesch Reading Ease (FRE)}
Llama 4 produced substantially more accessible justifications, achieving an FRE score of 51.6, which falls within the typical range for clinical documentation (50-70 \cite{wu2013applying,challener2025flesch}). In contrast, GPT-4o's score of 34.73 places its output in the ``difficult'' range, reflecting a clear prioritization of technical precision over linguistic simplicity. This highlights that readability is not an inherent model limitation but a highly tunable parameter. A simple directive within the prompt to target a specific readability level could readily align a model's output with the practical demands of a clinical workflow, making this a key consideration for system design.

\subsubsection{Flesch-Kincaid Grade Level (FKGL)}
Both models produced outputs with an FKGL appropriate for clinical documentation, which typically targets a Grade 8-12 standard \cite{wu2013applying,challener2025flesch}. Llama 4 scored an 8.5, equivalent to a mid-8th-grade reading level, while GPT-4o scored a 10.52, corresponding to an early high school complexity. These results confirm that both models can generate explanations that are linguistically accessible to their intended clinical audience, with Llama 4 producing slightly more straightforward text. This finding is crucial, as it demonstrates that the complexity of the underlying AI reasoning does not have to result in an equally complex explanation for the end-user.

\subsubsection{Clinical Grounding (CG)}
GPT-4o demonstrated superior factual grounding with a CG of 0.75, slightly outperforming Llama 4 at 0.72. Both models reliably reproduced most quantitative details from the input data, a critical capability in clinical domains where numerical precision is paramount. However, neither model achieved perfect accuracy. The persistence of minor hallucinations, even in state-of-the-art models, is a crucial finding that directly validates the necessity of a contestable framework. It underscores that human expertise is not merely beneficial but indispensable for verifying AI-generated claims before they can inform clinical decisions.

\subsubsection{Self-Correction Accuracy (SCA)}
Llama 4 demonstrated superior responsiveness to contestation, achieving a higher SCA score of 0.33 compared to GPT-4o's 0.17. This suggests the smaller model is more adaptive and willing to revise its initial assessment in light of contradictory evidence. In contrast, GPT-4o's lower score reflects a more conservative, risk-averse behavior, where it tends to default to the baseline model's prediction. This finding reveals a crucial trade-off: while larger models may offer greater factual grounding, smaller models might be more amenable to the corrective feedback that is central to a truly collaborative human-AI system.

Our results indicate that Llama 4 produced more readable and adaptive explanations, whereas GPT-4o demonstrated superior factual grounding at the cost of higher linguistic complexity. This reveals a critical design trade-off between factual veracity and adaptive reasoning. The selection of an LLM is therefore not a simple technical optimization but a decision that fundamentally shapes the nature of the human-AI partnership, balancing the need for a reliable adjudicator against that of a collaborative and correctable partner.

\begin{table}[t!]
    \centering
    \caption{Human-centric evaluation metrics. 
    The arrows ($\uparrow$/$\downarrow$) indicate whether higher or lower values are better.}
    \begin{tabularx}{\linewidth}{l *{4}{>{\centering\arraybackslash}X}}
    \toprule
    \textbf{Model} & 
    \textbf{FRE $\uparrow$} & 
    \textbf{FKGL $\downarrow$} & 
    \textbf{CG $\uparrow$} & 
    \textbf{SCA $\uparrow$} \\
    \midrule
    llama-4-scout-instruct (17B) & 51.60 & 8.50 & 0.72 & 0.33 \\ gpt-4o (200B) & 34.73 & 10.52 & 0.75 & 0.17 \\

    \bottomrule
    \end{tabularx}
    \label{tab:human-metrics}
\end{table}

\section{Discussion}
Our work demonstrates the feasibility of Motion2Meaning, a framework that successfully integrates AI-powered gait analysis with a contestable, human-in-the-loop interface. The findings confirm that it is possible to build systems that provide objective insights into motor symptoms while preserving essential clinical oversight through a structured and auditable workflow. We structure our discussion around two key themes. First, we reflect on the potential of AI-powered wearable gait analysis to transform PD care, considering our predictive model's performance and limitations. Second, we analyze the challenges and future directions for developing truly human-centered CAI, drawing specific insights from the performance of our XMED safeguard and LLM-based interaction components.

\subsection{The Potential of AI-Powered Wearable Gait Analysis in PD Care}
Our results, showing that a 1D-CNN can effectively classify disease severity from raw gait signals, reinforce the potential of wearable sensors to shift patient assessment from episodic clinical snapshots to continuous, longitudinal monitoring. This objective data stream offers a powerful complement to subjective tools like the UPDRS, enabling clinicians to more precisely track therapy response, detect subtle motor fluctuations, and quantify changes in fall risk. The ultimate promise of this technology is a move toward more personalized and proactive treatment adjustments that can tangibly improve patients' quality of life.

However, a purely gait-focused approach has inherent limitations. The nuances of intermediate disease stages and the multi-system nature of PD, which includes significant non-motor symptoms like cognitive impairment and sleep disturbances, underscore that gait is only one piece of a complex clinical picture. Therefore, the true value of this technology lies in augmenting, not replacing, holistic clinical judgment. The critical next step is to move beyond a unimodal biomarker toward a comprehensive digital phenotype of PD. Future work should focus on multi-modal fusion, integrating gait data with other streams like speech analysis and sleep tracking. This approach is essential not only for improving predictive accuracy but for capturing the true syndromic nature of the disease, leading to more robust and clinically relevant AI models.

\subsection{Toward Human-Centered Contestable AI in Healthcare}
The fallibility of our predictive model, despite its reasonable performance, underscores the critical need for robust human-in-the-loop systems. Our work addresses this by creating a multi-layered verification framework where the XMED module acts as an automated safeguard and the LLM serves as an interactive adjudicator. This design directly confronts the fundamental challenge of balancing automation's efficiency with the necessity of human oversight. However, our findings also illuminate several key challenges that must be addressed to advance the development of truly effective and collaborative CAI.

\subsubsection{Technical and Methodological Limitations}
Our evaluation confirms that explanation discrepancies can effectively signal prediction errors, validating XMED's potential as an automated safeguard. The scope of this validation, however, has two key considerations. First, the XMED flagging mechanism relied on an empirically derived discrepancy threshold. While effective for this study, its optimal calibration for broader clinical use is a natural next step for future work. Second, our framework was developed and validated exclusively within the context of PD, which served as a motivating use case. Its applicability and potential modifications for other diagnostic domains remain an important area for further exploration.

Furthermore, our experiments with LLMs highlight an interesting performance trade-off between the more adaptive reasoning of Llama 4 and the more factually grounded but conservative style of GPT-4o. This suggests that the selection of a model for adjudication involves balancing different priorities. A crucial open question, therefore, is whether this trade-off is inherent to LLM-based adjudication or is an artifact of using general-purpose models. The next logical step is to systematically evaluate models specifically fine-tuned on medical corpora, such as the Med-PaLM \cite{Singhal2023,singhal2025toward} and BioMistral \cite{labrak2024biomistral} families, and to explore the capabilities of next-generation generalist models like GPT-5 \cite{wang2025capabilities}, to determine if they can resolve the tension between adaptability and reliability. The modular design of our framework is a key strength in this regard, as it readily allows for the substitution of the predictive model, the LLM adjudicator, or the underlying dataset. This flexibility ensures that the core principles of our contestable system can be adapted and refined as new models and new clinical use cases emerge.

\subsubsection{Challenges in Clinical Validation}
A primary challenge arising from our study is the need for a comprehensive clinical evaluation. While our technical validation provides encouraging proof-of-concept results, the true measure of our framework's success lies in its real-world utility. This necessitates a formal clinician-in-the-loop pilot study to move beyond automated metrics and assess the system's impact on actual clinical practice. Such a study would involve close collaboration with neurologists to evaluate the correctness, clarity, and actionability of the LLM's outputs and to measure key workflow metrics like review time and contest rates. Gathering this contextual, human-centered evidence through detailed case studies is the essential next step for responsibly translating this research prototype into a validated clinical tool.

\subsubsection{Future Direction on Evaluation Metrics}
A truly human-centered approach requires a paradigm shift in evaluation. This involves conducting formal pilot studies with expert clinicians to move beyond accuracy to a suite of utility metrics. These can be divided into automated assessments of the AI's output and observational measures of user interaction. For instance, an automated metric like Clinical Terminology Grounding (CTG) could assess if an explanation is grounded in professional language by calculating the percentage of sentences containing terms from a predefined clinical lexicon. A high score would indicate reasoning more plausible to an expert. Observational metrics could quantify interaction efficiency through measures like Time to Decision (TTD), the duration until a user validates or contests a finding, and Interaction Length (IL), the number of conversational turns needed to reach that decision. Beyond these, our evaluation could be extended by adapting established metrics of user reliance, such as the Relative AI Reliance (RAIR), Relative Self-Reliance (RSR), and Appropriateness of Reliance (AoR) proposed in \cite{10.1145/3581641.3584066}, as well as the broader range of measures collected in \cite{app12199423,hsiao2021roadmap}.
\section{Conclusion}
This work demonstrates that Motion2Meaning unites SOTA AI gait analysis with a contestable, human-centered framework to deliver objective, accountable, and clinically viable interpretations of Parkinsonian motor symptoms. By combining accurate 1D-CNN classification with the XMED safeguard for uncertainty detection and the LLM-driven ``contest and justify'' workflow, the system ensures that clinical expertise remains central to decision making. Beyond advancing PD care, our results highlight a broader design principle: high-stakes medical AI must not only explain but also enable contestation, creating systems that are transparent, auditable, and aligned with regulatory requirements. Looking forward, integrating multimodal data, expanding to diverse populations, and tailoring domain-specific LLMs will further strengthen this approach. As global frameworks increasingly mandate explainability and contestability, Motion2Meaning provides a concrete step toward trustworthy AI that augments rather than replaces human judgment, setting a foundation for safer and more responsible deployment of AI in healthcare.

\bibliographystyle{splncs04}
\bibliography{ref}

\end{document}